# Reduce The Wastage of Data During Movement in Data Warehouse

Ahmed Mateen
Department of Computer Science,
University of Agriculture Faisalabad,
Pakistan

Lareab Chaudhary
Department of Computer Science,
University of Agriculture Faisalabad,
Pakistan

## ABSTRACT
In this research paper so as to handle Data in warehousing as well as reduce the wastage of data and provide a better results which takes more and more turn into a focal point of the data source business. Data warehousing and on-line analytical processing (OLAP) are vital fundamentals of resolution hold, which has more and more become a focal point of the database manufacturing. Lots of marketable yield and services be at the present accessible, and the entire primary database management organization vendor nowadays have contributions in the area assessment hold up spaces some quite dissimilar necessities on record technology compare to conventional on-line transaction giving out application. This article gives a general idea of data warehousing and OLAP technologies, with the highlighting on top of their latest necessities. So tools which is used for extract, clean-up and load information into back end of a information warehouse; multidimensional data model usual of OLAP; front end client tools for querying and data analysis; server extension for proficient query processing; and tools for data managing and for administration the warehouse. In adding to survey the circumstances of the art, this article also identify a number of capable research issue, a few which are interrelated to data wastage troubles. In this paper use some new techniques to reduce the wastage of data, provide better results. In this paper take some values, put in anova table and give results through graphs which shows performance.

## Keywords
Data warehouse, IO, OLAP, analytical, Hadoop, volume, velocity, variety, function, Big Data.

## 1. INTRODUCTION
Data warehouse provide a new and wide idea of the company gives better performance of data base, a enhanced organization and attainment of the data, where the creation of a data daunting task especially the helpful warehouse schema description. In fact, the search for technique of modeling data warehouses has become path blooming. In the warehouse is a research center of attention on the volume, velocity and variety which is used for to handle a large amount of data [1].

### 1.1 Peta-scale data
Peta-scale data high performance procedure which is generally to a great extent constructed with heterogeneous CPU( central processing unit) as well as GPU nodes to increase high performance and throughput of data [2] [3].

### 1.2 Map reduce
MapReduce is a structure for capably handing out the investigation of big data on a huge amount of servers[4]. It was residential for the back end of Google's exploration to facilitate a big amount of goods servers to proficiently procedure the study of vast records of web pages composed from the entire above the globe [5] [6] [7].

### 1.3 Hadoop
Apache Hadoop software collection is a structure that allow for the distributed dispensation of big data sets crosswise cluster of computer with easy encoding model[8][9]. Hadoop was originally encouraged by article's available by Google, exactness its advance to handle an sudden large amount of information, and has since turn out to be the standard for storing, giving out and analyze hundreds times of data in amount of terabytes, and even petabytes of figures [10].

## 2. PREVIOUS WORK
The data in data warehouse not completely send to the other server because a large amount of data and its miss the least priority data and always prefer to the highest priority data snow use new techniques which works efficiently and provide better result [11][12][13]. In this paper it is illustrate that how to overcome on peta-scale data and how to reduce the data wastage during moving data from one server to other and how get better results. so use different techniques and get better result and efficiency.

## 3. MATERIAL AND METHODS
In this paper it's illustrate the new techniques which is used for reduce the wastage of data. By using this overcome the wastage of data and also used some values to shows the results.

### 3.1 New techniques
$E_{node2ssd}$ = PSSD .occupied •N • (λa + λc) $BW_{host2ssd}$• tsim

Thus, the vitality devoured by the information examination is:

Eactivessd = PSSDbusy •N•λa•( 1/BWfm2c + 1/BWc2m+ 1/TSSD_k )•tsim

After information examination assignments, both checkpoint and investigation information are composed to the PFS by N S SSDs. Each SSD composes S • (α • λa + λc) measure of information at a transmission capacity of S •BWPFS/N[14].

**Essd2pfs = PSSD occupied • N2 • (α • λa + λc) S • BWPFS • tsim**

At long last, the unmoving vitality utilization can be computed utilizing the assessed all out unmoving time, using the bustling time gauge above:

**Eidlessd = PSSD unmoving • (N/S• tsim −Eactivessd + Essd2pfs/PSSDbusy)**

This is on the grounds that both CPUs and SSDs are included in the information exchange, which is not a piece of tsim, and does not perform information era. Strikingly, a speedier I/O transmission capacity to the SSD may decrease the I/O time at the figure hubs (right now the essential inspiration to convey SSDs in HPC frameworks.

**Tiosaving = N•(λa+λc)/BWPFS − S•(λa+λc)/BWSSD.**

Subsequently, the vitality reserve funds at all process hubs because of this can be composed as:

**Eiosaving = N • Tiosaving • Pserver unmoving**

In the wake of considering the greater part of the segments, vitality utilization at all SSDs can be communicated as takes after:

**ESSD = Enode2ssd + Eactivessd + Essd2pfs + Eidlessd − Eiosaving**

Execution and Energy Modeling for Offline Processing: (by an element of α) examination information as indicated by their offer of the entire information, i.e., N•λa/M , accepting M hubs are utilized to perform disconnected from the net investigation. So also, each of theM hubs need to prepare the information at the preparing rate, Terverk , for a given portion k. Again it accept that every hub gets the suitable offer of the PFS transfer speed 1 N • BWPFS.

## 3.2 Calculation

New Data Warehouse Designing Approach in view of Principal part Analysis, called DWDARPA, gets as info every one of the information set[7]. It yields the elements outlining the most related variables, from which the information distribution center composition will be created [15]. Truth be told, DWDARPA, an iterative procedure, works in four phases: The primary stage compresses the information for the most logical variables and studies the connection between's these variables by ascertaining the relationship lattice. At the second stage, continue the extraction of components in light of the connected variables[16]. At this level, compute the aggregate fluctuation mirroring the level of data that is the element including every one of the variables. At long last, the information distribution center blueprint is produced and can be accepted by the area master.

Data – Xi…….n

Result =C :c component

Start:

Check the variable relationship using correlation

Corr= Computed value

Store Corr

Retrieving Factors

For (m =1 to<n by 1)

V= Commutative variance

Var (m,0) = Variance Computed

Store Var

Identify C and V

Return

Exit

## 3.2 Analysis

In this study there are showing how to use the datastore and mapreduce functions to process a large amount of file-based data movement and it analysis. The MapReduce algorithm is a mainstay of many modern "big data" applications. This study operates on a single computer, but the code can scale up to use Hadoop.

Throughout this example, the data set is a collection of records for Data warehouse between 1987 and 2008. A small subset of the data set is also included with MATLAB® to allow us to run this and other examples without the entire data set.

Creating a datastore allows us to access a collection of data in a chunk-based manner. A datastore can process arbitrarily large amounts of data, and the data can even be spread across multiple files. it can create a dataset for a group of tabular content library, a SQL database (Database Toolbox™ required) or a Hadoop® Distributed File System (HDFS™).

ds = datastore('Filename');

dsPreview = preview(ds);

dsPreview(:,10:15)

The datastore automatically parses the input data and makes a best guess as to the type of data in each column. In this case, utilize the 'TreatAsMissing' Name-Value couple argument to have datastore replace the missing ethics correctly. For numeric variables (such as 'AirTime'), datastore replaces every 'NA' string with a NaN value, which is the IEEE arithmetic representation for Not-a-Number.

## 3.3 Scan for Rows of Interest

Datastore objects contain an internal pointer to keep track of which chunk of data the read function returns next. Use the has data and read functions to step through the entire data set, and filter the data set to only the rows of interest. In this case, the rows of interest are server on Data warehouse.

## 3.4 Use mapreduce to perform a computation

A use of mapreduce is to find the longest flight time in the entire airline data set. To do this:

1. The "mapper" function computes the maximum of each chunk from the datastore.

2. The "reducer" function then computes the highest value amongst the entire of maxima computed by the calls to the mapper function.

First, reset the datastore and filter the variables to the one column of interest.

reset(ds);

ds.SelectedVariableNames = {'ActualElapsedTime'};

Once the mapper and reducer functions are written and saved in your current folder, you can call mapreduce using the datastore, mapper function, and reducer function. If you have Parallel Computing Toolbox (PCT), MATLAB will automatically start a pool and parallelize execution. Use the readall function to display the results of the MapReduce algorithm.

result = mapreduce(ds, @maxTimeMapper, @maxTimeReducer);

readall(result)

### 3.5 Mapreduce progress
Map   0% Reduce   0%

Map 50% Reduce   0%

Map 100% Reduce   0%

Map 100% Reduce 100%

**Table 1: Data of various Servers**

| Server Num | Tail Num | Actual Elapsed Time | CRS Elapsed Time | Extra Time | Delay |
|---|---|---|---|---|---|
| 1503 | 'NA' | 53 | 57 | 'NA' | 8 |
| 1550 | 'NA' | 63 | 56 | 'NA' | 8 |
| 1589 | 'NA' | 83 | 82 | 'NA' | 21 |
| 1655 | 'NA' | 59 | 58 | 'NA' | 13 |
| 1702 | 'NA' | 77 | 72 | 'NA' | 4 |
| 1729 | 'NA' | 61 | 65 | 'NA' | 59 |
| 1763 | 'NA' | 84 | 79 | 'NA' | 3 |
| 1800 | 'NA' | 155 | 143 | 'NA' | 11 |

## 4. RESULTS
**Model 2:** DW= $\alpha + OIS\beta_1 + PIS\beta_2 + TIS\beta_3 + \varepsilon$

**Table 2: Regression Analysis with Model 2**

| Summary output | |
|---|---|
| **Regression Statistics** | |
| Multiple R | 0.479899 |
| R Square | 0.230303 |
| Adjusted R Square | -0.15455 |
| Standard Error | 3.253204 |
| Observations | 10 |

**Table 3: Model 2**

| ANOVA | | | | | |
|---|---|---|---|---|---|
| | df | SS | MS | F | Significance F |
| Regression | 3 | 19 | 6.333333 | 0.598425197 | 0.6391055 |
| Residual | 6 | 63.5 | 10.58333 | | |
| Total | 9 | 82.5 | | | |

**Model 3:** DW= $\alpha + DQ\beta_1 + SQ\beta_2 + \varepsilon$

**Table 4: Regression Analysis with Model3**

| Summary output | |
|---|---|
| Multiple R | 0.426401 |
| R Square | 0.181818 |
| Adjusted R Square | -0.05195 |
| Standard Error | 3.105295 |
| Observations | 10 |

**Table 5: Model 3**

| ANOVA | | | | | |
|---|---|---|---|---|---|
| | df | SS | MS | F | Significance F |
| Regression | 2 | 15 | 7.5 | 0.777777778 | 0.495421 |
| Residual | 7 | 67.5 | 9.642857 | | |
| Total | 9 | 82.5 | | | |

**Table 6: Data sending and receiving Delay**

| Unique Carrier | Server Num | Sending Delay | Receiving Delay | Origin |
|---|---|---|---|---|
| S1 | 121 | -9 | 0 | C1 |
| S2 | 1021 | -9 | -1 | C2 |
| S3 | 519 | 15 | 8 | C3 |
| S4 | 254 | 9 | 8 | C4 |
| S5 | 701 | -17 | 0 | C5 |
| S6 | 673 | -9 | -1 | C6 |
| S7 | 91 | -3 | 2 | C7 |
| S8 | 335 | 18 | 4 | C8 |
| S9 | 1429 | 1 | -2 | C9 |
| S10 | 53 | 52 | 13 | C10 |

In Model 2 Data wastage (DW) is dependent variable and Organization Implementation Success (OIS), Project Implementation Success (PIS) and Technical Implementation Success (TIS) are independent variables. After applying regression analysis on Model 2 have found Data wastage is almost 48% dependent on above said independent variables as Multiple R is 0.479899. The F calculated is 0.599 and F table value is 0.7256 it means it cannot accept that data wastage is not dependent upon said independent factors.

In Model 3 Data wastage (DW) is dependent variable and Data Quality (DQ) and System Quality (SQ) are independent variables. After applying regression analysis on Model 3 it have found Data wastage is nearly 43% dependent on above said independent variables as Multiple R is 0.426401. The F calculated is 0.78 and F table value is 0.4945 it means it cannot reject that data wastage is dependent upon said independent factors.

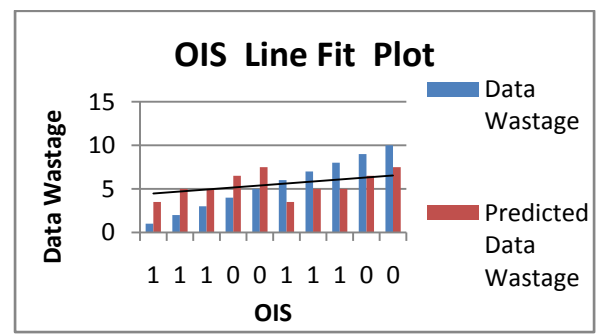

**Fig. 1: Data wastage and organization implementation success**

Above figure shows that Data wastage is less dependent upon Organization Implementation Success as line is touching fewer points.

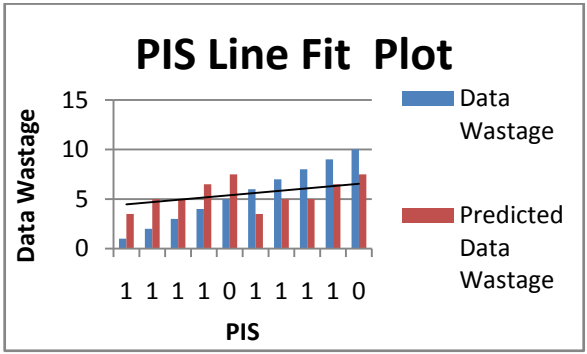

**Fig. 2: Data Wastage and Project Implemntation Success**

Above figure shows that Data wastage is less dependent upon Project Implementation Success as line is touching fewer points.

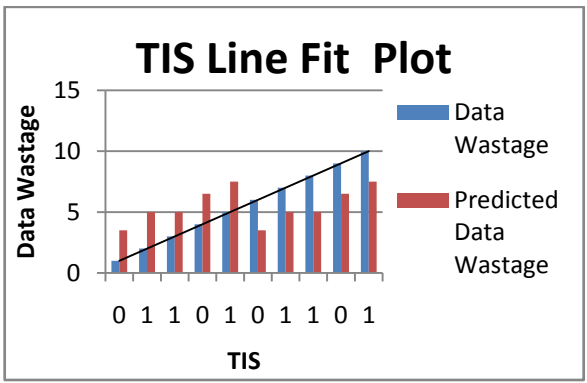

**Fig. 3: Data wastage and technical implementation success**

Above figure shows that Data wastage is more dependent upon Technical Implementation Success as line is touching more points. Multiple R suggested that implementation success factors up to 48% critical for data wastage. The factor Technical Implementation Success is showing more influence than other two factors.

The behavior of the mapper function in this application is more complex. For every server found in the input data, use the `add` function to add a vector of values. This vector is a count of the number of client for that carrier on each day in the 21+ years of data. The code is the key for this vector of values. This ensures that all of the data for each carrier will be grouped together when `map reduce` passes it to the reducer function.Save the following mapper function (`Server Mapper`).

Parallel map reduce execution on the parallel pool:

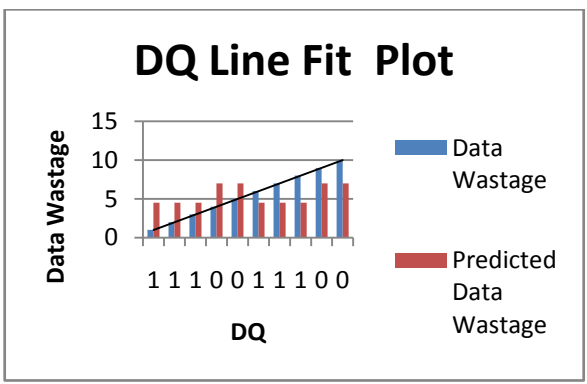

**Fig. 4: Data wastage and data quality**

Above figure shows that Data wastage is more dependent upon Data Quality as line is touching many points.

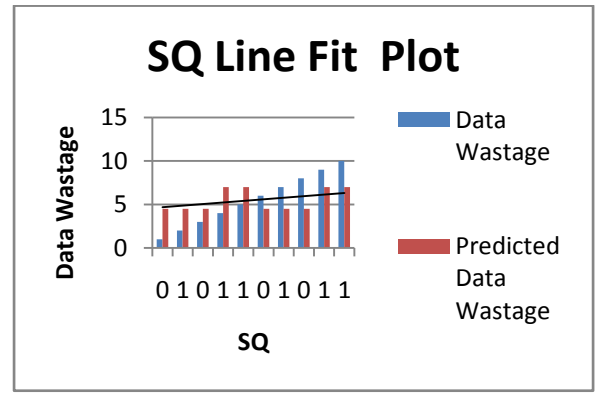

**Fig. 5: Data wastage and system quality**

Above figure shows that Data wastage is less dependent upon System Quality as line is touching fewer points. Multiple R is showing that data and system are 43% impact on data wastage factor. Data quality is better line fit then System quality, means data wastage is more dependent upon data quality than system quality.

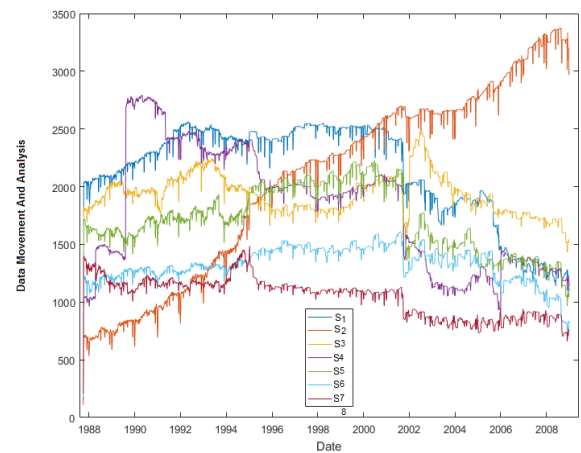

**Fig. 6: Data movement comparision**

Map  0% Reduce   0%

Map 50% Reduce   0%

Map 100% Reduce   0%

Map 100% Reduce 100%

These result shows in graphs and tell about their performance who data wastage reduce and how many data are loss during movement. In future a new approach will be introduced with the joint mixture of above these models with statically analyzed approach which will be definitely help full in cloud data handling

## 5.  CONCLUSION AND FUTURE WORK
In this research it is concluded that get better results to using some new techniques and get efficient and better result. There are also used ANOVAs table with some values and get results. These result shows in graphs and tell about their performance who data wastage reduce and how many data are loss during movement. In the data store automatically parses the input data and makes a best guess as to the type of data in each column and scan of rows insert data sending and receiving dely. Also used some table values which show the data wastage and movement of data around server and results which shows in tabular form and data movement comparison. It shows the Data Wastage and Project Implemntation Success, Data Wastage and Technical Implementation Success, Data Wastage and Data Quality, Data Wastage and System QualityIn future a new approach will be introduced with the joint mixture of above these models with generic algorithm of Distributed Cloud Algorithm which will be definitely helpful and provide a road way for handling the cloud data